\documentclass[aps,pra,onecolumn,groupedaddress,showpacs,amsmath,amssymb]{revtex4-1}

\usepackage{ifthen}		

\usepackage{graphics}
\usepackage{graphicx}
\usepackage{amsmath}
\usepackage{amssymb}
\usepackage{bm,color}

\newcommand{\be}{\begin{eqnarray}}
\newcommand{\ee}{\end{eqnarray}}

\renewcommand{\theequation}{\arabic{equation}}

\DeclareMathAlphabet{\mathpzc}{OT1}{pzc}{m}{it}

\begin{document}

\title{Generalized lattice Wilson--Dirac fermions in ($1+1$) dimensions 
for atomic quantum simulation and topological phases} 

\author{Yoshihito Kuno$^{1}$, Ikuo Ichinose$^{2}$, and Yoshiro Takahashi$^{1}$}

\affiliation{Department of Physics, Graduate School of Science, Kyoto University, Kyoto 606-8502, Japan}
\affiliation{Department of Applied Physics, Nagoya Institute of Technology, Nagoya, 466-8555 Japan}

\begin{abstract}
The Dirac fermion is an important fundamental particle appearing in high-energy
physics and topological insulator physics.
In particular, a Dirac fermion in a one-dimensional lattice system exhibits
the essential properties of topological physics.
However, the system has not been quantum simulated in experiments yet. 
Herein, we propose a one-dimensional generalized lattice 
Wilson-Dirac fermion model and study its topological phase structure.
We show the experimental setups of an atomic quantum simulator for the model,
in which two parallel optical lattices with the same tilt for
trapping cold fermion atoms and  a laser-assisted hopping scheme are used.
Interestingly, we find that the model exhibits nontrivial topological phases characterized 
by gapless edge modes and a finite winding number
in the broad regime of the parameter space.
Some of the phase diagrams closely resemble those of the Haldane model.
We also discuss topological charge pumping and a lattice Gross-Neveu model 
in the system of generalized Wilson-Dirac fermions.  
\end{abstract}
\maketitle

\section*{Introduction} 

The quantum simulation \cite{Nori} of Dirac fermions is of fundamental importance 
because they are ubiquitous in theoretical physics.
Dirac fermions appear in high-energy physics \cite{Ryder,Rothe} 
and the condensed matter physics, e.g. topological matter \cite{Wen},
graphene physics, etc.
In recent years, topological phases have become one of the most interesting 
subjects in physics, where Dirac fermions play an important key role \cite{Shen, Asboth}. 
In particular, a variety of one-dimensional (1D) lattice models 
have been extensively studied 
from the view point of nontrivial topological phases \cite{Zhu,Deng,Lang, Matsuda,Xu,Ganeshan,Hu}.
Experiments on cold atomic gases in an optical lattice have started 
to construct a ``quantum simulator" of 1D topological models.
Very recently, the experimental realization of a lattice topological model has been
reported in Ref. \cite{Jo}.
As one of the recent remarkable successes in experiments
concerning 1D topological models, we note 
the realization of topological Thouless pumping \cite{Nakajima,Lohse} 
and a ladder topological model in a synthetic dimensional optical lattice \cite{Mancini}.

Despite such experimental successes, the Dirac fermion model on a lattice called the
Wilson-Dirac model \cite{Wilson} is still a toy model in the sense that
it has not been realized and not yet quantum simulated in experiments. 
The 1D Wilson-Dirac model is the simplest and fundamental model that exhibits the
essence of a topological insulator \cite{Shen}. 
Thus, it is important to propose a quantum simulation for it and 
investigate its topological properties. 
Herein, we introduce a 1D generalized Wilson-Dirac model (GWDM) 
as an important quantum simulator.
We propose feasible experimental setups for the 1D GWDM and 
investigate the phase diagram of the 1D GWDM theoretically, 
in particular, the locations of topological phases. 

Schemes for realizing quantum simulators for the standard Dirac-fermion systems,
using cold-atomic gases in continuum and lattice systems, have been already proposed.
Some of them are a Raman coupling scheme \cite{Galitskii,Ruostekoski,1Dtopological,Bermudez}, 
a modulation method on a tilted lattice \cite{Garreau}, and 
an effective model of two-component cold atoms in a 1D optical 
superlattice \cite{Cirac}. 
The above works focus on the standard Dirac fermions.
In this work, we are interested in constructing a quantum simulator for extended
lattice Wilson-Dirac fermions, which include the ordinary 
Wilson-Dirac fermions on the lattice
as a specific case and has a large parameter space to be realized by experiments.
The 1D GWDM, which contains nontrivial phases in the hopping terms, 
is an interesting model by itself because these phases work as free parameters 
that change the physical properties of the ordinary Wilson-Dirac fermion model, 
e.g. the symmetries of the Hamiltonian, the energy spectrum, the ground state including
nontrivial topological phases, etc.
Actually, in the experimental setups for the 1D GWDM, the phases can be controlled 
by a laser-assisted hopping scheme, which is familiar in experiments on cold atomic 
systems. 

In order to realize the Dirac fermions by cold atomic gases in an optical lattice, 
the greatest difficulty is the creation of the Dirac-gamma matrices 
from the nearest-neighbor (NN) hopping amplitudes of cold atoms.
To this end, we use two different internal states of a fermionic atom 
and two parallel ``tilted'' optical lattices. 
This setup is an important platform for realizing the 1D GWDM.
In particular, we explain a general scheme for the generation of  Dirac-gamma
matrices by using a laser-assisted hopping technique. 
Furthermore, to understand the general construction scheme, 
we propose a concrete set up by using ${}^{171}$Yb atoms.
After that, we study the symmetry properties and the topological phases 
of the 1D GWDM and provide the expected ground-state phase diagrams. 
Finally, we put $\hbar=1$ throughout this paper.

\section*{\textbf{Results}}
\paragraph*{\textbf{Generalized Wilson-Dirac fermions}}\label{SecII}
As explained in the introduction,
we consider two internal states for a single fermion and denote them by 
$\Psi_{j}=(a_{j},b_{j})^t$ at lattice site $j$. 
The GWDM in a 1D spatial lattice is defined by the following Hamiltonian, 
\begin{eqnarray}
H^{(g)}_{\rm GWDM}&=&\sum_{j}\Psi^{\dagger}_{j}\Gamma_{z}(\Delta)\Psi_{j}\nonumber\\
&-&\sum_{j}\biggr[\Psi^{\dagger}_{j+1}\Gamma^{h}_{z}(\theta_{a},\theta_{b})\Psi_{j}+\mbox{h.c.}\biggr] \nonumber\\
&+&\sum_{j}\biggl[\Psi^{\dagger}_{j+1}\Gamma_{x}(\theta^{+},\theta^{-})\Psi_{j}+
\Psi_{j+1}\Gamma^{*}_{x}(\theta^{+},\theta^{-})\Psi^{\dagger}_{j}\biggr],  \nonumber\\
\label{generalModel}
\end{eqnarray}
with
\begin{eqnarray}
\Gamma_{z}(\Delta)&=&
\begin{bmatrix}
  \Delta & 0 \\
  0 & -\Delta
\end{bmatrix}
=\Delta\sigma_{z},\label{Gammaz}\\
\Gamma^{h}_{z}(\theta_{a},\theta_{b})&=&
\begin{bmatrix}
  |J_{a}|e^{i\theta_{a}} &0 \\
  0 & |J_{b}|e^{i\theta_{b}}
\end{bmatrix},\label{Gammahz}\\
\Gamma_{x}(\theta^{+},\theta^{-})&=&
\begin{bmatrix}
  0 & |J^{-}_{ab}|e^{i\theta^{-}} \\
  |J^{+}_{ab}|e^{-i\theta^{+}} & 0
\end{bmatrix},\label{Gammax}\\
\Gamma^{*}_{x}(\theta^{+},\theta^{-})&=&
\begin{bmatrix}
  0 & |J^{-}_{ab}|e^{-i\theta^{-}} \\
  |J^{+}_{ab}|e^{i\theta^{+}} & 0
\end{bmatrix},\label{Gammaxc}
\end{eqnarray}
where $\sigma_{z}$ is the Pauli matrix, $\theta_{a}$, $\theta_{b}$, $\theta^{+}$, 
and $\theta^{-}$ are site-independent phases, 
$\Delta$ is an energy-offset; and $|J_{a}|$, $|J_{b}|$, $|J_{ab}|$ and $|J^{-}_{ab}|$ 
are hopping amplitudes. 
The different internal states $a_j$ and $b_j$ originate from, e.g. an internal spin,
and have different energy levels. 
In this case, the energy splitting is nothing but a hyperfine energy splitting, 
which can be created by the Zeeman effect in an external magnetic field. 
 
We express the model in Eq.(\ref{generalModel}) 
in terms of the fermion creation and annihilation operators 
$a^{\dagger}_{j} (a_{j})$ and $b^{\dagger}_{j} (b_{j})$, as
\begin{eqnarray}
H^{(g)}_{\rm GWDM}&=&H_{\rm spinOL}+H_{\rm ahop}+H_{\rm bhop}+H^{+}_{ab}+H^{-}_{ab},
\\ 
H_{\rm spinOL}&=&\sum_{j=1}\Delta( a^{\dagger}_{j}a_{j}-b^{\dagger}_{j}b_{j}),
 \label{H_spinOL}\\
H_{\rm ahop}&=&-\sum_{j}J_{a}a^{\dagger}_{j+1}a_{j}+\mbox{h.c.},\label{Hahop}\\
H_{\rm bhop}&=&-\sum_{j}J_{b}b^{\dagger}_{j+1}b_{j}+\mbox{h.c.},\label{Hbhop}\\
H^{+}_{ab}&=&\sum_{j}J^{+}_{ab}a^{\dagger}_{j}b_{j+1}+\mbox{h.c.}\label{Habhop+}\\
H^{-}_{ab}&=&\sum_{j}J^{-}_{ab}a^{\dagger}_{j}b_{j-1}+\mbox{h.c.},\label{Habhop-}
\end{eqnarray}
where $J_{a}=|J_{a}|e^{i\theta_{a}}$, $J_{b}=|J_{b}|e^{i\theta_{b}}$, 
$J^{+}_{ab}=|J_{ab}|e^{i\theta^{+}}$, and $J^{-}_{ab}=|J^{-}_{ab}|e^{i\theta^{-}}$. 
In the following section, we shall show feasible methods for constructing each term 
in the above Hamiltonian $H^{(g)}_{\rm GWDM}$ in experiments on ultra-cold 
fermion gases. 
Before detailing the theoretical proposal, we note that by setting 
the hopping amplitudes as $|J_{a}|=|J_{b}|=t$ and $|J^{+}_{ab}|=|J^{-}_{ab}|=t'$ and 
the phases as $\theta_{a}=0$, $\theta_{b}=\pi$ and $\theta^{+}=-\theta^{-}=-\pi/2$, 
the Hamiltonian in Eq.(\ref{generalModel}) reduces to the (1+1)D version of 
the ordinary Wilson-Dirac fermion model \cite{Wilson}, in which
the Dirac gamma matrices are given by $\gamma_{0}=\sigma_{z}$, $\gamma_{1}=\sigma_{y}$ 
and $\gamma_{5}=\gamma_{0}\gamma_{1}$, respectively.
It should be emphasized that the phase conditions, $\theta_{a}=0$, $\theta_{b}=\pi$ 
and $\theta^{+}=-\theta^{-}=-\pi/2$ can be realized in real experiments
by tuning the incident angle of Raman lasers. 
Hereafter, we call these conditions the {\it Dirac condition}. \\

\paragraph*{\textbf{Theoretical proposal for quantum simulation}}\label{SecIII}

Let us explain the general setup scheme 
for the Hamiltonian in Eq.(\ref{generalModel}) by ultra-cold atomic gases.
To this end, we use two different internal states of a single fermionic atom 
in an optical lattice. 
In particular, the most important problem is the creation of 
the generalized gamma matrices 
in the Hamiltonian $H^{(g)}_{\rm GWDM}$ given by Eqs.(\ref{Gammaz})-(\ref{Gammaxc}). 
Generally speaking, the experimental setup consists of three steps: 
(i) prepare two different internal states of a fermionic atom
and set two parallel deep optical lattices with the same tilt.
(ii) apply four types of laser-assisted hopping
that generate the matrices in Eqs.(\ref{Gammaz})-(\ref{Gammaxc}) 
by using some excitation lasers in addition to the off-resonant laser of the optical lattice,
and
(iii) tune the intensity and frequency of the excitation lasers and 
set the appropriate incident angle of the excitation lasers to realize the 
uniform phase condition. 

In order to clarify the above setup, we shall explain each step in detail in the rest
of this section. \\

\paragraph*{\textbf{Two parallel optical lattice}}\label{SecIIIA} 

\begin{figure}[t]
\begin{center} 
\includegraphics[width=12cm]{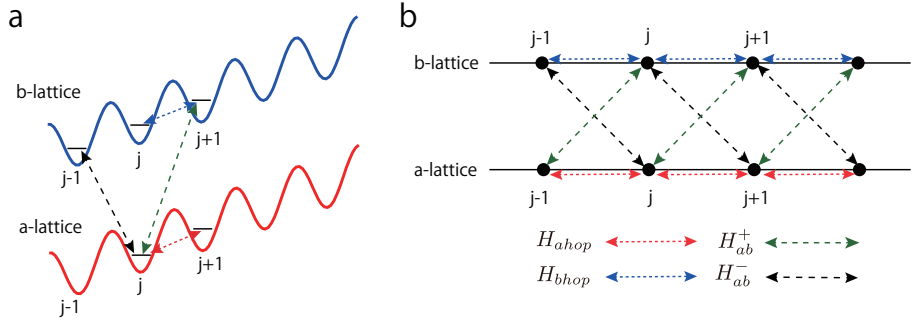}
\end{center} 
\caption{ (a) Two parallel optical lattice. Each lattice traps a different internal state of fermion. 
The two lattices have the same tilt. 
(b) Four types of hopping term. In the exchange hopping term denoted by the black and green dashed arrows, 
the fermionic atoms hop to a different site in a different optical lattice 
with changing the internal spin.}
\label{Setup}
\end{figure}

In our proposal, we first prepare the two internal states of the fermionic atom 
denoted by $|a\rangle$ and $|b\rangle$ and consider 
two 1D parallel optical lattices with the same tilt \cite{tilt_lattice,tilt_lattice2}.
Each optical lattice can trap one of two states, 
$|a\rangle$ or $|b\rangle$. 
Then, we set the optical lattices sufficiently deep to suppress 
the natural hopping process between NN lattice sites.  
Here, we call the optical lattice trapping the state $|a\rangle$ the ``a-lattice" 
and the other optical lattice trapping the state $|b\rangle$ the ``b-lattice".
We apply the tight-binding picture to each optical lattice system and assume that
the potential minimums of the two lattices exist at the same locations \cite{Mandel}. 
The lattice site label $j$ is used for the a- and b-lattices as shown in 
Fig.\ref{Setup} (a), 
i.e. the a- and b-lattices comprise a parallel optical lattice system. 
In this system, by choosing two appropriate internal levels of the fermionic atom
$|a\rangle$ and $|b\rangle$, 
an energy-offset $\Delta_{ab}$ at site $j$ can be generated. 
In the second-quantized tight-binding picture, the energy-offset $\Delta_{ab}$ 
leads to $\sum_{i}\frac{\Delta_{ab}}{2}(a^{\dagger}_{j}a_{j}-b^{\dagger}_{j}b_{j})$, 
where the tight-binding operators of $|a\rangle$ and $|b\rangle$ are regarded 
as the operators $a_{j}$ and $b_{j}$ defined in the previous section.
Therefore, the energy-offset part $H_{\rm spinOL}$ of Eq.(\ref{H_spinOL}) is 
identified as $\Delta = \Delta_{ab}/2$.\\

\paragraph*{\textbf{Four types of laser-assisted hopping}}\label{SecIIIC}
\begin{figure}[t]
\begin{center} 
\includegraphics[width=6cm]{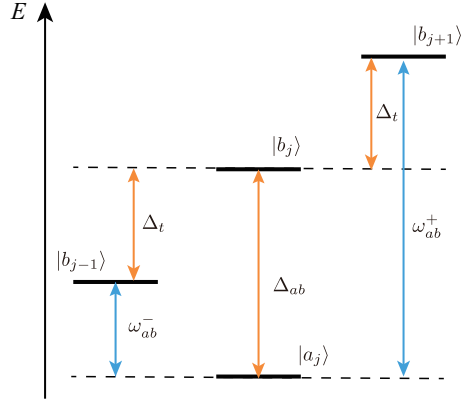}
\end{center} 
\caption{Energy condition for realizing the laser-assisted hopping
$\Lambda^{a_{j}}_{b_{j+1}}$ and $\Lambda^{a_{j}}_{b_{j-1}}$.}
\label{Setup2}
\end{figure}

For the realization of the hopping terms in Eqs.(\ref{Hahop})-(\ref{Habhop-}), 
we apply four types of laser-assisted hopping to the two parallel optical lattice system.
Laser-assisted hopping is created by the $\Lambda$-shaped scheme explained 
in Methods. 
By using the two parallel lattices, which are tilted by the same amount by 
certain experimental techniques, 
the tilted energy difference $\Delta_{t}$ between the NN lattice can be introduced.  
We denote the four types of hopping corresponding to 
Eqs.(\ref{Hahop})-(\ref{Habhop-}) by $\Lambda^{a_{j}}_{a_{j+1}}$, $\Lambda^{b_{j}}_{b_{j+1}}$, 
$\Lambda^{a_{j}}_{b_{j+1}}$ and $\Lambda^{a_{j}}_{b_{j-1}}$.
For example, the label on $\Lambda^{a_{j}}_{a_{j+1}}$ means the NN hopping between 
$a_{j}$ and $a_{j+1}$ on the $a$-lattice, 
which corresponds to the hopping term in Eq.(\ref{Hahop}) 
as shown in Fig.\ref{Setup} (a) and Fig.\ref{Setup} (b).
The other labels have the similar meanings. 

Next, to establish
$\Lambda^{a_{j}}_{a_{j+1}}$, $\Lambda^{b_{j}}_{b_{j+1}}$, $\Lambda^{a_{j}}_{b_{j+1}}$ and $\Lambda^{a_{j}}_{b_{j-1}}$ without interfering with each other, 
suitable tunings of the on-site energy-offset, the lattice tilt 
and the excitation laser frequencies used in the laser-assisted hopping amplitudes are required. 
One can directly create $\Lambda^{a_{j}}_{a_{j+1}}$ and $\Lambda^{b_{j}}_{b_{j+1}}$ 
by choosing an appropriate excited state for each state.
However, $\Lambda^{a_{j}}_{b_{j+1}}$ and $\Lambda^{a_{j}}_{b_{j-1}}$ have to be carefully 
prepared because we need to prohibit or highly suppress the Rabi coupling
$a^{\dagger}_{j}b_{j}+a_{j}b^{\dagger}_{j}$ at the same site. 
Furthermore, in the case where the two component fermionic atom 
originates from different $z$-components of the internal spin, 
we must select the appropriate polarization of the excitation lasers in creating $\Lambda^{a_{j}}_{b_{j+1}}$ and $\Lambda^{a_{j}}_{b_{j-1}}$. 
To satisfy these requirements, we show the general tuning condition as the energy
diagram shown in Fig.\ref{Setup2},
where,  $|a_{j}\rangle$, $|b_{j}\rangle$, $|b_{j+1}\rangle$, and $|b_{j-1}\rangle$ are 
two different internal states of an atom on lattice sites $j$, $j+1$ and $j-1$, respectively. 
The energy splitting $\Delta_{ab}$ between $|a_{j}\rangle$ and $|b_{j}\rangle$ 
is related to the on-site energy-offset $2\Delta=\Delta_{ab}$, as explained before.  
$\omega^{+}_{ab}=\Delta_{ab}+\Delta_{t}$ 
is the frequency difference of the two excitation lasers used in the laser-assisted 
hopping $\Lambda^{a_{j}}_{b_{j+1}}$. 
(For a detailed definition, see Methods.)
$\omega^{-}_{ab}=\Delta_{ab}-\Delta_{t}$ is the corresponding quantity of
$\Lambda^{a_{j}}_{b_{j-1}}$. 
To realize $\Lambda^{a_{j}}_{b_{j+1}}$ and $\Lambda^{a_{j}}_{b_{j-1}}$ independently, 
we need to impose the following condition: 
 $\Delta_{ab} > \Delta_{t} \gg \delta^{+(-)}_{b}$, where $\delta^{+(-)}_{b}$ is 
the two photon detuning 
used in the laser-assisted hopping $\Lambda^{a_{j}}_{b_{j+1}}$ ($\Lambda^{a_{j}}_{b_{j-1}}$). 
Actually, we take $\delta^{+(-)}_{b}$ to be zero to create a resonance between NN sites.
Then, in order to sufficiently separate the resonance conditions of $\Lambda^{a_{j}}_{b_{j+1}}$ and $\Lambda^{a_{j}}_{b_{j-1}}$,  
the difference between the two resonance denoted by $\omega^{+}_{ab}-\omega^{-}_{ab}=2\Delta_t$ 
should be large compared with the two-photon Rabi frequency in Eq.(\ref{hopbase}).
If these conditions are satisfied in the two parallel optical lattices, 
the transition probability of the same site is negligibly small 
(at least strongly suppressed ). 
Then, $\Lambda^{a_{j}}_{b_{j+1}}$ and $\Lambda^{a_{j}}_{b_{j-1}}$ are dominant. 

As explained above, by applying the four types of laser-assisted hopping
$\Lambda^{a_{j}}_{a_{j+1}}$, $\Lambda^{b_{j}}_{b_{j+1}}$, $\Lambda^{a_{j}}_{b_{j+1}}$ and $\Lambda^{a_{j}}_{b_{j-1}}$ 
to the two parallel optical lattices with the same tilt, 
we can design the hopping terms in Eqs.(\ref{Hahop})-(\ref{Habhop-}). 
Therefore, we can produce a quantum simulator of the 1D GWDM. 

Furthermore, controlling the parameters of the excitation lasers enables us 
to set the uniform phases $\theta_{a}$, $\theta_{b}$, $\theta^{+}$, and $\theta^{-}$ and the uniform hopping amplitudes rather freely. \\
\paragraph*{\textbf{Concrete example using ${}^{171}$Yb}}

\begin{figure}[t]
\begin{center} 
\includegraphics[width=16cm]{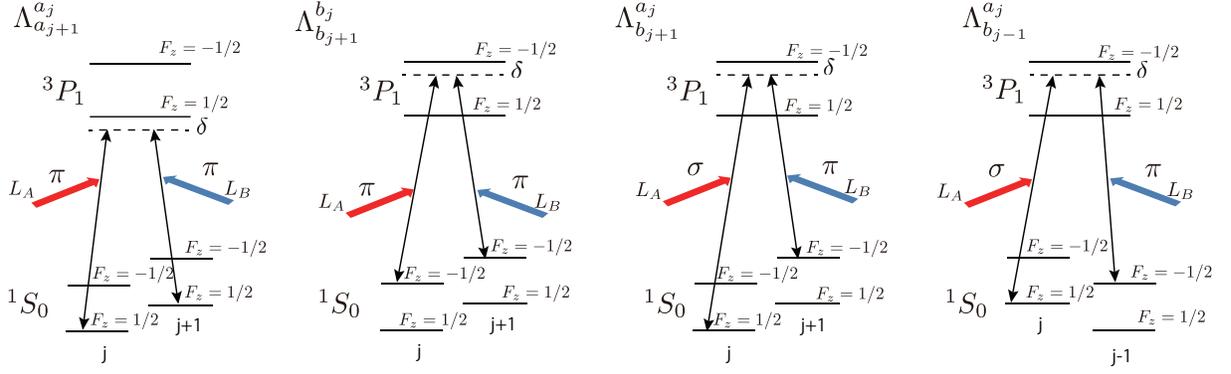}
\end{center} 
\caption{Schematics of four types of laser-assisted 
hopping in the ${}^{171}$Yb atom system.
The energy difference between the NN sites ($j$ and $j\pm1$), and that 
between $F_{z}=1/2$ and $F_{z}=-1/2$ in the ${}^{1}S_{0}$ manifold 
at the same site are 
$\pm\Delta_{t}$ and $\Delta_{ab}$, respectively. 
The detunings for the excited 
states take the same value $\delta$.}
\label{Setup_b}
\end{figure}

In general, the above theoretical proposal can be performed 
by using some atomic species, e.g. alkali atoms.
As one candidate, we consider ${}^{171}$Yb 
atoms. 
In particular,
we employ the two internal states of  ${}^{171}$Yb, $|{}^{1}S_{0}$, $F_{z}=1/2\rangle$ 
and $|{}^{1}S_{0}$, $F_{z}=-1/2\rangle$ 
as two component fermionic state, i.e. 
$$|{}^{1}S_{0}, F_{z}=1/2\rangle\to |a \rangle,  \;\;
|{}^{1}S_{0}, F_{z}=-1/2\rangle\to |b \rangle.$$
Then, the energy splitting $\Delta_{ab}$ can be generated and controlled by 
a uniform magnetic fields, i.e.
$\Delta_{ab}\to \Delta_{ab}(B_0)$. 
Actually, the nuclear g-factor of ${}^{171}$Yb is 0.985; therefore,
the value of $\Delta_{ab}$ is set to be 75 kHz with a magnetic field of 100 G. 
Furthermore, to create $\Lambda^{a_{j}}_{a_{j+1}}$, $\Lambda^{b_{j}}_{b_{j+1}}$, 
$\Lambda^{a_{j}}_{b_{j+1}}$ and $\Lambda^{a_{j}}_{b_{j-1}}$, 
we have to select the appropriate four sets of three states 
$\{|A\rangle,|B\rangle,|E\rangle\}$ in the $\Lambda$-shaped scheme. 
(See Methods.) 
Here, we show the selection of the ${}^{171}$Yb internal states 
in Table~\ref{171Yb_laser_asissted}. 
Here, the $(L_{A},L_{B})$ line in Table~\ref{171Yb_laser_asissted} expresses 
the pattern of polarization of the two excitation lasers. 
(The excitation lasers with $\pi$ or $\sigma^{\pm}$ polarization are considered here.) 
$\Lambda^{a_{j}}_{b_{j+1}}$ and $\Lambda^{a_{j}}_{b_{j-1}}$ can be controlled independently. 
Furthermore,  $\Lambda^{a_{j}}_{a_{j+1}}$ and $\Lambda^{b_{j}}_{b_{j+1}}$ can be 
independently controlled since the two excited states $|{}^{3}P_{1}$,
$F_{z}=1/2\rangle$
 and $|{}^{3}P_{1}$, $F_{z}=-1/2\rangle$ can be well separated 
on the order of 100MHz with magnetic field having
a reasonable strength about 100 G. 
Figure~\ref{Setup_b} shows schematics of the four types of 
laser-assisted hopping corresponding to Table~\ref{171Yb_laser_asissted}. 
The energy difference between the two excited states and 
the natural widths of the two excited states 
are denoted by $\omega_{z}$, $\Gamma_{E_{1}}$ and $\Gamma_{E_{2}}$, respectively.
Then, the detuning $\delta$ for each type of laser-assisted hopping is allowed to satisfy $\Gamma_{E_{1}},\Gamma_{E_{2}}\ll \delta\ll \omega_{z}$ 
since $\omega_{z}/(2\pi)=$ 100[MHz] and $\Gamma_{E_{1(2)}}/(2\pi)\sim$ 200[kHz]. 
This condition allows an independent laser assisted hopping scheme. 
The mutual interference could be suppressed to be on the order of $\delta/\omega_z\ll 0.1$. 
That is, the four types of laser-assisted hopping can be produced independently. 
Here, we comment that the overlap integral $\tilde{J}_{j,j+1}$, which is explicitly defined 
in Eq.(\ref{hopbase}) in Methods, depends on the shape 
and location of the potential minimum of the Wannier functions in the ${}^{3}P_{1}$ 
excited states, which are used in the four types of laser-assisted hopping.
Generally, to make $\tilde{J}_{j,j+1}$ have a finite and reasonably large value, 
the location of the potential minimum needs to be set on 
the potential maximum of the ${}^{1}S_{0}$ lattice \cite{Jaksch,Grimm}. 
Furthermore, the Wannier function of the excited states needs to be sufficiently
broad so that the overlap integral has a sufficiently large value. 
To this end, the relation between the polarization of the ${}^{3}P_{1}$ excited states,
denoted by $\alpha_{P}$, and that of ${}^{1}S_{0}$, denoted by $\alpha_{S}$, plays 
an important key role.
If $|\alpha_{P}|-|\alpha_{S}|<0$ and ${\rm sgn}(\alpha_{P})=-{\rm sgn}(\alpha_{S})$, 
the above requirement is satisfied. 
In fact, the
Yb atom satisfies the conditions for typical wave-lengths 
of optical lattice lasers, e.g. 532 nm and 1064 nm, etc \cite{Barker,Ovsyannikov}.  
Therefore, the overlap integral between the ${}^{1}S_{0}$ and 
${}^{3}P_{1}$ Wannier functions can be sufficiently large.\\

\begin{table}[t]
\begin{center} 
  \begin{tabular}{|c|c|c|c|c|} \hline
                       & $\Lambda^{a_{j}}_{a_{j+1}}$ & $\Lambda^{b_{j}}_{b_{j+1}}$ & $\Lambda^{a_{j}}_{b_{j+1}}$ & $\Lambda^{a_{j}}_{b_{j-1}}$  \\ \hline
    $|A\rangle$   & ${}^{1}S_{0}$, $F_{z}=1/2$ & ${}^{1}S_{0}$, $F_{z}=-1/2$ & ${}^{1}S_{0}$, $F_{z}=1/2$ &  ${}^{1}S_{0}$, $F_{z}=1/2$\\ \hline
    $|B\rangle$   & ${}^{1}S_{0}$, $F_{z}=1/2$ & ${}^{1}S_{0}$, $F_{z}=-1/2$ & ${}^{1}S_{0}$, $F_{z}=-1/2$ &  ${}^{1}S_{0}$, $F_{z}=-1/2$\\ \hline
    $|E\rangle$   & ${}^{3}P_{1}$, $F_{z}=1/2$ & ${}^{3}P_{1}$, $F_{z}=-1/2$ & ${}^{3}P_{1}$, $F_{z}=-1/2(1/2)$ &${}^{3}P_{1}$, $F_{z}=-1/2(1/2)$\\ \hline 
    $(L_{A},L_{B})$ & ($\pi$,$\pi$) & ($\pi$,$\pi$) & ($\sigma (\pi)$ ,$\pi(\sigma)$)  &  ($\sigma(\pi)$,$\pi(\sigma)$)\\ \hline
  \end{tabular}
\end{center} 
\caption{Four types of laser-assisted hopping by using the hyperfine structure of ${}^{171}$Yb}
\label{171Yb_laser_asissted}
\end{table}

\paragraph*{\textbf{Phase diagram and topological phase}}\label{SecIV}

As a next step, we study whether or not the GWDM has nontrivial topological phases.   
The system contains the uniform phases $\theta_a $, $\theta_b $, $\theta^{+}$,
and $\theta^{-}$; then, we shall clarify their parameter regime
corresponding to topological phases.  
In what follows, we regard $\theta_a $, $\theta_b $, $\theta^{+}$, and $\theta^{-}$ 
as free parameters. 
The above phases are fully tunable in real experiments; see Methods.

To discuss the above problem, we first study the symmetries of the GWDM 
by using the symmetry-classification scheme in Refs.~\cite{Kitaev,Ryu}.
The symmetries of the system depend on the phases $\{\theta\}$.
We shall also obtain the energy spectrum of the GWDM on a finite lattice 
with open boundary condition (OBC). 
Then, the spectrum is expected to exhibit zero-energy edge states 
in some parameter regime of the phases $\{\theta\}$.
The existence of the zero-energy edge states is a direct signal of a
nontrivial topological phase in the bulk system by the bulk-edge correspondence.

Hereafter for simplicity, we impose conditions for the hopping amplitudes in
Eqs.(\ref{Hahop})-(\ref{Habhop-}) such as 
$|J_{a}|=|J_{b}|=|J^{+}_{ab}|=|J^{-}_{ab}|=1$ as they do not change the physical results. 
The above condition again can be realized in certain experimental setups \cite{Miyake2}. 

We consider the system under periodic boundary condition, which 
preserves the discrete translational symmetry. 
We first focus on the symmetries of the bulk Hamiltonian and also
the bulk topological properties of the GWDM.  
The bulk-momentum Hamiltonian $H_{\rm bulk}(k)$
is obtained from Eq.~(\ref{generalModel}) as,
\begin{eqnarray}
&&H_{\rm bulk}(k)=
\begin{bmatrix}
  \Delta - 2\cos( k+\theta_{a}) & e^{-ik+i\theta^{+}}+e^{ik+i\theta^{-}} \\
  e^{-ik-i\theta^{-}}+e^{ik-i\theta^{+}} & -\Delta-2\cos( k+\theta_{b})
\end{bmatrix},
\label{generalModelF}
\end{eqnarray}
where we have taken the lattice spacing as the length unit.
Under the Dirac conditions $\theta_{a}=0$, $\theta_{b}=\pi$, 
and $\theta^{+}=-\theta^{-}=\pi/2$, 
$H_{\rm bulk}(k)$ is just the bulk-momentum Hamiltonian of 
the ordinary Wilson-Dirac fermion:
\begin{eqnarray}
H_{\rm bulk}(k)
= [\Delta - 2\cos k]\sigma_{z}+ [2\sin k]\sigma_{x}
\equiv \bf{d}(k) \cdot \bf{\sigma},
\label{bulk_Dirac}
\end{eqnarray}
where ${\bf{d}}(k)= (d_{x},d_{z}) =(2\sin k, \Delta - 2\cos k)$.
This is the base model of 1D topological insulator\cite{Shen} and 
belongs to the BDI class Hamiltonian. [See later discussion.] 
Then, the nontrivial topological phase can be characterized 
by the winding number $N_w$ \cite{Asboth}, 
which is obtained by integrating the vector trajectory of 
${\bf{d}}(k)$ defined by,
\begin{equation}
N_{w}=\int^{\pi}_{-\pi}\frac{dk}{2\pi}\frac{\bm{d}(k)}{|\bm{d}(k)|}\times \frac{d}{dk}\biggl( \frac{\bm{d}(k)}{|\bm{d}(k)|}\biggl).
\label{winding_number}
\end{equation}
For a nontrivial topological phase, $N_{w}=+1$ or $-1$, whereas for a trivial insulating 
phase $N_{w}=0$.
In the parameter regime $-2\leq \Delta\leq 2$, a nontrivial topological phase with $N_{w}=+1$ is known to exist \cite{Shen,Asboth}.  

We shall also consider the finite lattice system of the 1D GWDM with OBC 
later and show the existence of degenerate
zero-energy edge states by diagonalizing the Hamiltonian of the 1D lattice system 
with the system size $L=100$ (generally, we take $L$ to be an even integer). 
The zero-energy edge state is a direct signal of the existence of nontrivial 
topological phases.\\

\paragraph*{\textbf{Symmetries, topological phases and zero-energy edge modes 
in the 1D GWDM}}

\begin{figure}[h]
\begin{center} 
\includegraphics[width=10cm]{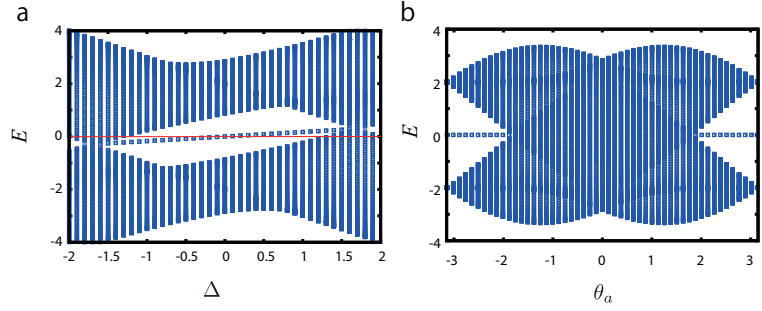}
\end{center} 
\caption{ (a) Energy spectra for $\theta_{a}=3\pi/4$ and $\theta_{b}=0$. 
(b) Energy spectra for $\Delta=0$ and $\theta_{b}=0$}
\label{CSspectrum}
\end{figure}

We shall show how to construct topologically nontrivial Hamiltonians in the 1D GWDM, 
and provide the experimental conditions for the laser setups to realize them.
As classification theory indicates \cite{Kitaev,Ryu}, a 1D model, which has 
nontrivial topological phases, belongs to the BDI or AIII class.
This means that the 1D model must at least possess chiral symmetry \cite{Ryu2}. 
The relevant symmetries are the time-reversal symmetry (${\cal T}$) and 
charge-conjugation
symmetry (${\cal C}$) for the classification scheme.
The system has time-reversal symmetry if and only if the Hamiltonian $H$
satisfies the following condition;
\begin{equation}
{\cal T} : \; U^\dagger_T {\cal K} H {\cal K} U_T =U^\dagger_TH^\ast U_T=H,
\label{timesym}
\end{equation}
where $U_T$ is a unitary operator and ${\cal K}$ is the complex-conjugation operator.
Similarly, for charge-conjugation symmetry (particle-hole symmetry),
\begin{equation}
{\cal C} : \; U^\dagger_C {\cal K} H {\cal K} U_C =U^\dagger_CH^\ast U_C=-H,
\label{chargesym}
\end{equation}
where $U_C$ is again a unitary operator.
The BDI class has ${\cal T}$ and ${\cal C}$ symmetries with    
${\cal T}^2=+1$ and ${\cal C}^2=+1$, whereas the AIII class has only
${\cal S}\equiv {\cal T}\cdot {\cal C}$ symmetry, which is called chiral symmetry.
Under ${\cal K}$, $k$ and $\theta$'s transform as 
$(k,\theta\mbox{'s}) \to -(k,\theta\mbox{'s})$.
Then, it is seen that the Hamiltonian of the ordinary Wilson-Dirac fermion
[Eq.~(\ref{bulk_Dirac})] has both time-reversal and charge-conjugation symmetries
with $U_T=\sigma_z$ and $U_C=\sigma_x$, respectively. 
From the time-reversal and charge conjugation operators, 
the chiral operator is directly obtained as $U_S=\sigma_{y}$. 

To search the parameter regime of the chiral symmetric Hamiltonian in the 1D GWDM, 
we first assume the Dirac condition, i.e. $\theta^{+}=-\pi/2$ and $\theta^{-}=\pi/2$,
but relax $\theta_{a}$ and $\theta_{b}$ as free parameters.
Then, we show the typical behavior of the energy spectra of the finite lattice system  
including the edge modes. 
In Fig.~\ref{CSspectrum} (a), we plot the energy spectra for $\theta_{a}=3\pi/4$ 
and $\theta_{b}=0$ by varying the parameter $\Delta$. 
The results show the spectrum of the edge modes is located at the center of the
spectra.
A close look at the calculations reveals that
the edge modes have non-vanishing energies, except for $\Delta=0$ and the spectrum
tilts along $\Delta$. 
This indicates that the present system does not have chiral symmetry, 
except in the case of $\Delta=0$.

To study further, by fixing $\theta_{b}=0$ and $\Delta=0$, we calculate
energy spectra by varying the parameter $\theta_{a}$. 
The results are shown in Fig.~\ref{CSspectrum}(b).
We find interesting behavior in the regimes of $-\pi\leq \theta_{a}\leq -\pi/2$ 
and $\pi/2\leq \theta_{a}\leq \pi$, i.e.
the zero-energy edge modes survive as long as the bulk-gap does not close. 

It is instructive to visualize the energy spectrum of the bulk system obtained 
from the bulk Hamiltonian $H_{\rm bulk}(k)$ in Eq.~(\ref{generalModelF}). 
For arbitrary $\theta_{a}$ and $\theta_{b}$ with $\Delta=0$, the bulk energy 
spectrum $E_{\pm}(k)$ is obtained as,
\begin{eqnarray}
E_{\pm}(k)&=&-(\cos (k+\theta_{a})+\cos (k+\theta_{b}))\nonumber\\
&&\pm\biggl[\biggl (\cos(k+\theta_{a})+\cos(k+\theta_{b})\biggr)^{2}\nonumber\\
&&-4\cos (k+\theta_{a})\cos (k+\theta_{b})-4\sin^{2}k \biggr]^{\frac{1}{2}}.
\label{bulk_energy}
\end{eqnarray}
When the first term on the right-hand side (RHS) of Eq.~(\ref{bulk_energy}) 
is non-vanishing, 
the spectrum is asymmetric, i.e. $E_{+}(k)\neq -E_{-}(k)$, this means that
the spectrum is non-relativistic. 
On the other hand,
once the first term vanishes, the spectrum is symmetric around $E=0$, i.e. 
$E_{+}(k)= -E_{-}(k)$. 
The spectrum is the relativistic (massive) Dirac type.
From this consideration, in order to make the 1D GWDM chiral symmetric, 
we should impose a condition such as 
\begin{eqnarray}
 \theta_{a}=\theta_{b}\pm\pi.
\label{Chiral}
\end{eqnarray} 
Hereafter, we call Eq.(\ref{Chiral}) the chiral symmetry (CS) condition. 
This observation of the bulk energy spectra gives an important insight about
the parameter regime of the topologically nontrivial Hamiltonian in the 
GWDM as well as the above numerical results of the edge modes in the finite system. 

\begin{figure}[t]
\begin{center} 
\includegraphics[width=10cm]{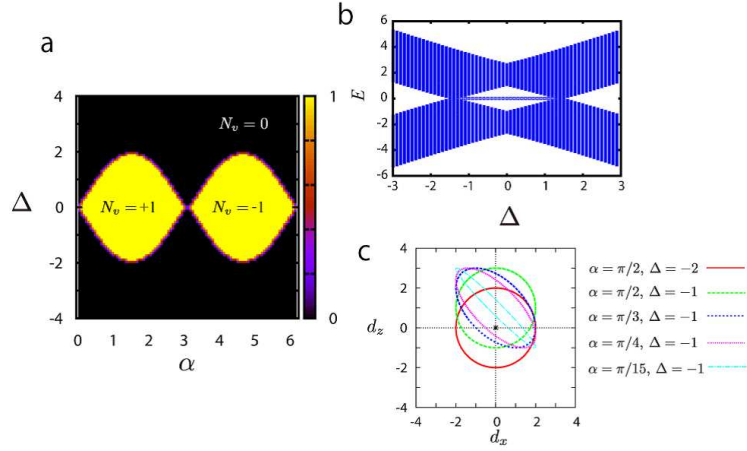}
\end{center} 
\caption{(a) Phase diagram of the 1D GWDM 
for $\theta^{a}=\theta^{b}\pm \pi$,  $\theta^{+}=-\theta^{-}$, and $\alpha\equiv\theta^{-}-\theta^{a}$.
The phase diagram has a similar structure to that
of the Haldane model on a honeycomb lattice.
$\Delta$ and $\alpha$ are free parameters.
(b) Energy spectra with a zero-energy edge state at $\alpha=\pi/4$.
(c) Hamiltonian trajectories when sweeping $k$. 
$\theta^{a}=\theta^{b}\pm \pi$, $\theta^{+}=-\theta^{-}$, and
$\alpha\equiv\theta^{-}-\theta^{a}$. $(d_{x},d_{z})=(0,0)$ is the gap closing point in our model.}
\label{SP5}
\end{figure}

Let us focus our attention on the CS case of the bulk Hamiltonian by setting 
$\theta_{a}=\theta_{b}\pm\pi$ in Eq.(\ref{generalModelF});
\begin{eqnarray}
H^{\rm CS}_{\rm bulk}(k)
&&= \Big[\Delta - 2\cos( k+\theta_{a})\Big]\sigma_{z}\nonumber \\
&&+\Big[\cos(-k+\theta^+)+\cos(k+\theta^-)\Big]\sigma_x \nonumber \\
&&+\Big[\sin(-k+\theta^+)+\sin(k+\theta^-)\Big]\sigma_y.
\label{HCS}
\end{eqnarray}
As the Hamiltonian $H^{\rm CS}_{\rm bulk}(k)$ in Eq.(\ref{HCS}) contains
all three components of the Pauli matrices, one may think that it cannot be chiral
symmetric unless further conditions are imposed.
However, we shall show that it is not only chiral symmetric but also 
time-reversal and charge-conjugate symmetric.
To this end, we introduce the rotated Pauli matrix
$\tilde{\sigma}_j(\rho)$ defined as follows (see Methods):
\begin{eqnarray}
\tilde{\sigma}_j(\rho) &\equiv& \exp(-i{\rho \over 2}\sigma_i)\sigma_j 
\exp(i{\rho \over 2}\sigma_i) \nonumber \\
&=&\sigma_j\cos\rho+\epsilon_{ijk}\sigma_k\sin\rho,
\label{rotatedPauli}
\end{eqnarray}
where $\epsilon_{ijk}$ is the totally anti-symmetric tensor, i.e. $\epsilon_{xyz}=1$, etc.
By using the rotated Pauli matrix $\tilde{\sigma}_x(\rho)$, it can be shown that
$H^{\rm CS}_{\rm bulk}(k)$ is expressed as, 
\begin{eqnarray}
H^{\rm CS}_{\rm bulk}(k)&&=
[\Delta - 2\cos( k+\theta_{a})]\sigma_{z}\nonumber \\
&&+\biggr[2\cos \biggr( k-\frac{\theta^{+}-\theta^{-}}{2}\biggl)\biggl]
\tilde{\sigma}_{x}\Big({\theta^{+}+\theta^{-}\over 2}\Big).
\label{HCS2}
\end{eqnarray}
This expression shows that the system Hamiltonian $H^{\rm CS}_{\rm bulk}(k)$
possesses time-reversal and charge-conjugation symmetries. 
In fact, for time-reversal symmetry, 
\begin{equation}
U_T=\exp[i(\theta^++\theta^-)\sigma_z],
\end{equation}
and for charge-conjugation symmetry,
\begin{equation}
U_C=\sigma_y\exp[i(\theta^++\theta^-)\sigma_z].
\end{equation}
We note that from the above consideration, the CS condition is an important 
condition for the BDI class bulk-momentum Hamiltonian.
That is, the CS condition in Eq.(\ref{Chiral}) is a sufficient condition for
the BDI class in our quantum simulator of the 1D GWDM. 
This means that {\it we do not need to implement the ordinary 1D Wilson-Dirac
fermion in the atomic simulator to simulate the topological properties of 
the Dirac model}.

We turn to the investigation of the phase diagram including nontrivial topological
phases under the CS condition. 
It is expected that interesting results are obtained 
because the additional phase parameters enlarge the regime of  topological phases 
from that of the standard Wilson-Dirac model.

By shifting the wave vector as $k\to k+(\theta^+-\theta^-)/2$, the Hamiltonian
$H^{\rm CS}_{\rm bulk}(k)$ is expressed as
\begin{eqnarray}
H^{\rm CS}_{\rm bulk}(k)&=&
\biggr[\Delta - 2\cos\biggl( k+\theta_{a}+\frac{\theta^{+}-\theta^{-}}{2}\biggr)\biggl]
\sigma_{z}\nonumber\\
&&+[2\cos k]\tilde{\sigma}_{x}((\theta^{+}+\theta^{-})/2).
\label{CSonly}
\end{eqnarray} 
Then, the Bloch vector is given by the following general form with an angle $\alpha$:
$$
{\bf d}(k)=(d_{x}(k),d_{z}(k))\equiv( 2\cos k,\Delta - 2\cos( k-\alpha)),
$$
where $\alpha=-\theta_{a}-(\theta^{+}-\theta^{-})/2$ in the present case.
We calculate the energy spectrum of the 1D GWDM on the finite lattice.
In particular, we focus on the zero-energy edge modes.
By diagonalizing the system Hamiltonian, we obtain the phase diagram including 
nontrivial topological phases in the $(\alpha$-$\Delta)$ plane.
We have used the existence of the zero-energy edge modes to identify
topological phases.  
\begin{figure}[t]
\begin{center} 
\includegraphics[width=10cm]{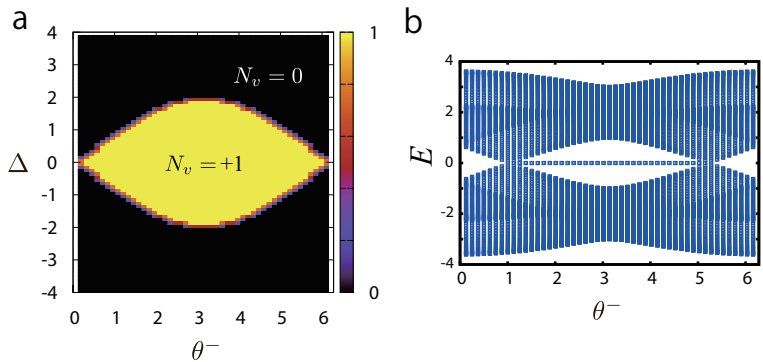}
\end{center} 
\caption{(a)Phase diagram of the existence of the zero-energy edge state 
for $\theta^{a}=\theta^{b}\pm \pi=0$ and $\theta^{+}=0$. 
$\theta^{-}$ and $\Delta$ are free parameters. 
(b) Energy spectra for $\Delta=1$}
\label{SP7}
\end{figure}
The obtained phase diagram for $\theta^+=-\theta^-$ is shown in 
Fig.\ref{SP5}(a) and a typical energy spectrum 
in the finite lattice system is shown in Fig.\ref{SP5}(b). 
In this case, the rotated Pauli matrix reduces to the original one, i.e. 
$\tilde{\sigma}_x((\theta^++\theta^-)/2) \to \sigma_x$.
As expected, there exist two topologically nontrivial phases, and 
they are labeled by the winding number $N_{w}=\pm 1$.  
Interestingly, the obtained phase diagram is similar to that of 
the Haldane model \cite{LLi,Haldane}.
Analytically, the phase boundaries between the trivial ($N_{v}=0$) and 
nontrivial topological phases ($N_{w}=\pm1$) 
are given by $\Delta=\pm \sin\alpha$. 
Compared to the Haldane model, 
the present topological phases are characterized by $N_{w}$
and not the Chern number, 
whereas $\alpha$ seems to correspond to the ``flux parameter" of the
Haldane model. 
At $\alpha=\pi/2$ or $\alpha=3\pi/2$, if $\theta_{a}=0$, the 1D GWDM 
reduces to the ordinary 1D Wilson-Dirac model.
The typical trajectories of 
${\bf d}(k)=(d_{x}(k),d_{z}(k))\equiv( 2\cos k,\Delta - 2\cos( k-\alpha))$ 
obtained by sweeping $k$ are plotted in Fig.\ref{SP5}(c), which gives 
the winding number $N_{w}$,
and in the Boch vector space, $(d_{x},d_{z})=(0,0)$ corresponds to the gap closing point.
From this plot, we can obtain the winding number $N_{w}$ [Eq.(\ref{winding_number})] 
from the bulk momentum Hamiltonian.

It is interesting to see the phase diagrams corresponding to the ``nontrivial" case
with the rotated Pauli matrix $\tilde{\sigma}_x(\theta)$
in Eq.(\ref{HCS}) [Eq.(\ref{CSonly})].
To this end, we fix $\theta_{a}=0$, $\theta_{b}=\pi$ and $\theta^{+}=0$; then,
the remaining parameters are $\theta^{-}$ and $\Delta$. 
The obtained phase diagram is shown in Fig.\ref{SP7} (a). 
The 1D GWDM on the finite lattice has a topological phase diagram including a broad
regime of nontrivial topological phase with $N_w=+1$, and there exist  
clear edge modes as seen in Fig.\ref{SP7}(b). 
From the results in Figs.\ref{SP5} and \ref{SP7}, we conclude that if the CS condition 
Eq.(\ref{Chiral}) is satisfied, nontrivial topological phases form in
rather broad parameter regimes.
This fact exhibits flexibility for the actual experimental realization of the 1D GWDM
as a quantum simulator of a 1D topological insulator. \\


\paragraph*{\textbf{Topological charge pumping and the realization of the 1D lattice 
Gross-Neveu model}}\label{SecV}

A topological pump can be realized in the 1D GWDM by adding a CS breaking term.
As an example, the 1D GWDM, which satisfies the CS condition in Eq.(\ref{Chiral}) 
and also $\theta^+=\theta^-$, can be a topological charge-pump model
by adding a $\sigma_{y}$-channel term to the GWDM.
Explicitly, the $\sigma_{y}$-channel term associated with 
$\tilde{\sigma}_{y}(\theta^{+})$ is given by   
\begin{eqnarray}
H_{\sigma_{y}}=M\sum_{j}\Psi_j^\dagger \tilde{\sigma}_{y}(\theta^{+})\Psi_j,
\label{ychannel}
\end{eqnarray}
where $M$ is the coupling constant of the $\sigma_{y}$ channel.
In experiments, this term can be created by using another laser-assisted hopping
scheme, as shown in Methods. 
With the term in Eq.(\ref{ychannel}), the bulk-momentum Hamiltonian of Eq.(\ref{HCS2}) is changed to
\begin{eqnarray}
H^{\rm P}_{\rm bulk}(k) 
&=& [\Delta - 2\cos (k+\theta_{a})]\sigma_{z}+ [2\sin k]\tilde{\sigma}_{x}(\theta^{+})
\nonumber \\
&&+ M \tilde{\sigma}_{y}(\theta^{+}).
\label{CPM}
\end{eqnarray}
As we vary the parameters $\Delta$ and $M$ adiabatically with the period $T$,
such as  
$$(\Delta,M)\to (\Delta\cos(2\pi t/T),M\cos(2\pi t/T))$$ 
[here, $T\gg 1$ for the adiabatic condition], then
the model in Eq.(\ref{CPM}) is expected to exhibit topological charge pumping phenomena.
The phenomena can be observed by measuring the bulk-particle current, 
which corresponds to a shift in the center of the Wannier function at an optical
lattice site \cite{Nakajima,Lohse}. 
A similar argument can be applied to  a more general case of the 1DGWDM.

It is interesting to include the interactions between atoms, in particular, 
those between the different internal states. 
The interspecies interactions such as, 
$$\sum_{j}V a^{\dagger}_{j}a_{j}b^{\dagger}_{j}b_{j}$$ with a coupling constant $V$
can be expressed in terms of spinor notation $\Psi_j$ as,
\begin{eqnarray}
V_{\rm int}&=&\sum_{j}V a^{\dagger}_{j}a_{j}b^{\dagger}_{j}b_{j} = \sum_{j}\frac{V}{2}(\Psi^{\dagger}_{j}\gamma_{0}\Psi_{j})^{2}.
\label{GNmodel}
\end{eqnarray}
Then, the model $H_{\rm WDM}+V_{\rm int}$ is nothing but the lattice version of 
the Gross-Neveu model \cite{Gross-Neveu}, which plays an important role
in quantum field theory and elementary particle physics.
Even in $(1+1)$D, the Gross-Neveu model has a nontrivial phase diagram
with a phase transition. 
A similar model  to the above has been proposed in Ref.\cite{Cirac} 
by using an optical superlattice. 
In real experiments, 
${}^{173}$Yb atom, for example, is a candidate, which has finite s-wave 
scattering length between the two different internal states in ${}^{1}S_{0}$, 
whereas the ${}^{171}$Yb atom has a much smaller on-site interaction.
Although adding the interaction term disturbs the conditions of lasers in 
laser-assisted hopping, the fine-tuning of lasers may allow one to 
realize a quantum simulator of the Gross-Neveu model. 

\section*{Discussion}\label{SecVI}

In this work, we theoretically proposed the realization of the 1D  generalized 
Wilson-Dirac Hamiltonian in a tilted optical lattice.
A combination of two parallel optical lattices with 
the same tilt and laser-assisted hopping
is employed for the atomic quantum simulation of the system.
As a concrete example, we suggested ${}^{171}$Yb fermionic atom and also
the candidates of energy levels to be used in laser-assisted hopping.
The model can be a quantum simulator of a 1D topological insulator.

Next, we studied the GWDM from the view point of symmetry classification theory,
which plays an important role in searching for topologically nontrivial phases.
Interestingly enough, we found that the CS condition is a sufficient condition 
that makes the 1D GWDM belong to the BDI class, and we verified this observation 
by numerically calculating the energy spectra and winding number. 
This result is important as it shows the flexibility and versatility of the 1D GWDM,
i.e. we do not need to create the exact 1D Wilson-Dirac model in experiments
as long as we focus on constructing a quantum simulator of a 
1D topological insulator.

We obtained the phase diagrams of the model including nontrivial topological phases,
and found that some of them have a feature similar to that of the Haldane model. 

Finally, we showed that the 1D GWDM possibly exhibits the topological charge 
pumping if the rotated $\sigma_{y}$-channel is included in this model.
We also suggested that by adding inter-species interactions, the model can be 
a quantum simulator of the lattice version of Gross-Neveu model \cite{Gross-Neveu}.
Analysis of the 1D GWDM with many-body interactions is an important subject 
and is expected to lead to richer nontrivial phases.
We hope that the proposal in this work will be used for the realization of 
atomic quantum simulators of 1D Dirac fermion physics for observing, e.g. the Zitterbewegung
phenomena in lattice systems \cite{Garreau,Vaishnav,Merkl,Qu,Leblanc}, 
and other related models \cite{Bermudez,Gholizadeh}.

\section*{Methods}

\paragraph*{\textbf{Laser-assisted hopping: General case}}\label{SecIIIB}
\begin{figure}[t]
\begin{center} 
\includegraphics[width=4cm]{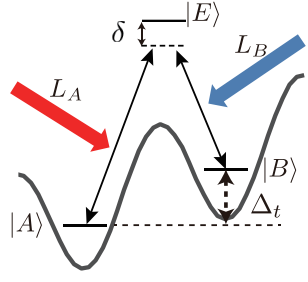}
\end{center} 
\caption{$\Lambda$-shaped schema} 
\label{Setup_assist}
\end{figure}
To generate the hopping terms in Eqs.(\ref{Hahop})-(\ref{Habhop-}) in 
the 1D GWDM,
we use excitation lasers in addition to the optical lattice lasers and generate 
laser-assisted hopping 
\cite{Jaksch, Aidelsburger,Miyake1,Miyake2,Gerbier,Dalibard,Lin,Goldman,Mancini,Celi}. 
This method is the standard method to create NN hoppings with a nontrivial phase. 
In general, three states with different energy levels are considered; then,
laser-assisted hopping is generated by using $\Lambda$-shaped scheme 
through Rabi coupling \cite{Goldman}.
Here, we explain the single $\Lambda$-shaped scheme proposed in
Refs.\cite{Miyake1,Keilmann,Jaksch,Greschner}.  

First, as shown in Fig.\ref{Setup_assist}, we consider two quantum states 
with different energy levels and different positions denoted by 
$|A\rangle$ and $|B\rangle$, and 
one excited state $|E\rangle$.
The energy gap between $|A\rangle$ and $|B\rangle$ is denoted by $\omega_{AB}$, 
and the energy gaps between $|A\rangle$ and $|E\rangle$, 
and between $|B\rangle$ and 
$|E\rangle$ are denoted by $\omega_{AE}$ and $\omega_{BE}$, respectively.  
Then by using two excitation lasers $L_{A}$ and $L_{B}$, we can couple $|A\rangle$ and $|B\rangle$ to $|E\rangle$. 
Here, $L_{A(B)}$ is set at the detuned-frequency $\omega_{AE(BE)}-\delta$, where $\delta$ is the detuning with $\delta\ll \omega_{AE(BE)}$ 
and $\delta \gg \Gamma_{E}$, where $\Gamma_{E}$ is the natural width of $|E\rangle$,   
and has the wave vector ${\bf k}_{A(B)}$, which is determined by $|{\bf k}_{A(B)}|=(\omega_{AE(BE)}-\delta)/c$ ($c$ is the speed of light).   
From the two excitation lasers, Rabi coupling can be generated through an electric dipole interaction. 
The Rabi couplings are denoted by $\Omega_{AE}$ and $\Omega_{BE}$.
In this setup, we can estimate the effects of the excited state $|E\rangle$ by 
using the second-order perturbation analysis. 
Consequently, the coupling between $|A\rangle$ and $|B\rangle$ is effectively generated. 
In the single particle picture, the coupling constant  between $|A\rangle$ and $|B\rangle$ 
in the rotating frame is given by $\frac{\Omega'_{AE} \Omega'_{BE}}{4\delta}$. 
A detailed calculation is shown in Supplementary Materials. 

Next, the single $\Lambda$-shaped scheme is applied to a 1D tilted deep single 
optical lattice, 
and we consider laser-assisted hopping.
The lattice tilt and deep lattice-depth suppress the natural tunneling between 
NN lattice sites.
The lattice tilt can be engineered, e.g. by using a magnetic field gradient, an
electric field (light-shift) gradient and gravity, and
leads to an energy difference $\Delta_{t}$ between each pair of NN lattice sites.
Then, the application of $L_{A}$ and $L_{B}$ to the entire system triggers 
a $\Lambda$-shaped transition of each NN lattice sites. 
Therefore, if we put non-interacting atoms in the 1D lattice, 
the tight-binding model is effectively given by 
\begin{eqnarray}
H^{2nd(g)}&=&\sum_{j}(\tilde{J}_{j,j+1}g_{j+1}^{\dagger}g_{j}+\mbox{h.c.}), \label{H2nd}\\
\tilde{J}_{j,j+1}&=&\frac{|\Omega'_{AE}||\Omega'_{BE}|}{4\delta}\nonumber\\
&&\times\int d{\bf r}\: W^{*}({\bf r}-{\bf r}_{j+1})e^{\delta{\bf k}\cdot {\bf r}}W({\bf r}-{\bf r}_{j}),
\label{hopbase}
\end{eqnarray}
where $g^{\dagger}_{j}(g_{j})$ is a creation (annihilation) operator of an atom on
lattice site $j$, and $\tilde{J}_{j,j+1}$ 
is a complex hopping parameter determined by a localized wave function 
$W({\bf r})\equiv w^{ws}(x)w(y)w(z)$. 
Here, $w^{ws}(x)$ is the Wannier-Stark state \cite{Gluck,Greschner}, determined 
by the tilted optical lattice and, 
$w(y)$ and $w(z)$ are the Wannier states, determined by the $y$- and $z$-direction 
optical lattices, 
which create a strong confinement potential creating a 1D system. 
$\delta{\bf k}$ is defined as $\delta{\bf k}={\bf k}_{A}-{\bf k}_{B}$. 
By appropriate tuning of the incident angles of the excitation lasers, $\delta{\bf k}$ 
can be uniform along the 1D lattice.
Here, it is noted that in Eq.(\ref{hopbase}), 
if we set $\omega_{BE}-\omega_{AE} \sim \Delta_{t}$, the tilt energy difference $\Delta_{t}$ 
between NN sites does not appear owing to the rotating wave approximation 
(RWA) with the rotating frame of $\omega_{AB}$ \cite{Miyake2}.
The hopping terms in Eq.(\ref{H2nd}) are a basic ingredient
for the creation of the hopping terms in Eqs.(\ref{Hahop})-(\ref{Habhop-}).\\

\paragraph*{Uniform phase creation}
\begin{figure}[h]
\begin{center}
\includegraphics[width=10cm]{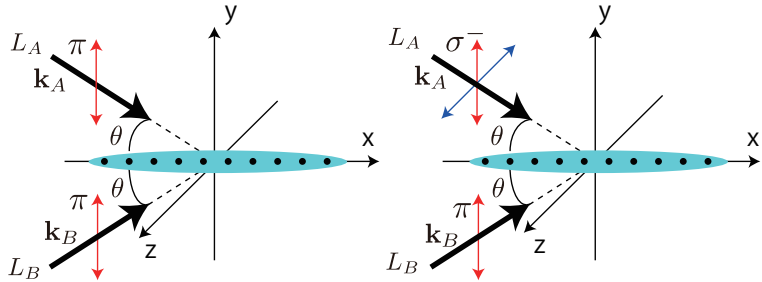}
\end{center}
\caption{Schematics of the incident lasers for laser-assisted hopping. 
The quantization axis is in the $x$-direction.}\label{SP8}
\end{figure} 
The phase created when in applying laser-assisted hopping is spatially dependent 
since the phase is determined by $\delta{\bf k}$ as in Eq.(\ref{hopbase}).
However, since our target model is 1D, 
if we prepare a three dimensional cubic optical lattice, 
1D optical lattice chains with uniform phases are created 
by making the remaining lattice potential sufficiently deep to confine atoms 
with many 1D tubes. 
When the direction of 1D tube in this lattice configuration is regarded as the
$x$-direction, the
condition $\delta{\bf k}=(0,k^{y}_{A}-k^{y}_{B},k^{z}_{A}-k^{z}_{B})$ leads to 
a uniform phase along the $x$-direction, 
even though the value of the uniform phase of each tube is different. 
Figure~\ref{SP8} shows schematics of incident lasers 
for the laser-assisted hopping with the uniform phase. 
The blue ellipses represent a 1D gas trapped in two parallel optical lattice. 
The left panel shows two types of laser-assisted hopping, $\Lambda^{a_{j}}_{a_{j+1}}$ and $\Lambda^{b_{j}}_{b_{j+1}}$. 
Similarly, the right panel shows two types of laser-assisted hopping, $\Lambda^{a_{j}}_{b_{j+1}}$ 
and $\Lambda^{a_{j}}_{b_{j-1}}$.
Both cases create the hopping with the uniform phase along the $x$-direction.\\

\paragraph*{Rotational transformed Pauli matrix}
The Pauli matrix can be transformed by performing a rotational transformation 
in the spin space. 
The full rotation of the spin space is determined by two rotational angles. 
In general, the rotated Pauli matrix $\tilde{\sigma}_{j}$ along the $i$-component 
spin ($i=1(x),2(y),3(z)$) axis 
is given by a formula incorporating the rotational angle $\rho$:
\begin{eqnarray}
\tilde{\sigma}_{j}(\rho)\equiv e^{-i\rho\sigma_{i}/2}\sigma_{j}e^{i\rho\sigma_{i}/2}
=\sigma_{j}\cos\rho+\epsilon_{ijk}\sigma_{k}\sin\rho. 
\end{eqnarray}
If one takes $(i,j,k)=(3,1,2)$, and sets $\rho=\phi$, the rotated $x$- and 
$y$-component Pauli matrices rotated around the $z$-spin axis are given as  
\begin{eqnarray}
\tilde{\sigma}_{x}(\phi)&\equiv&
\begin{bmatrix}
  0 & e^{i\phi} \\
  e^{-i\phi} & 0
\end{bmatrix},\label{rzsigmax}\\
\tilde{\sigma}_{y}(\phi)&\equiv&
\begin{bmatrix}  
  0 & -ie^{i\phi} \\
  ie^{-i\phi} & 0
\end{bmatrix}.\label{rzsigmay}
\end{eqnarray}
The rotated sigma matrices ($\tilde{\sigma}_{x}(\phi)$, $\tilde{\sigma}_{y}(\phi)$,$\sigma_{z}$) 
also satisfy the SU(2) commutation relation.
By the complex conjugate transformation ${\cal K}$,
\begin{eqnarray}
{\cal K}\tilde{\sigma}_x(\phi) {\cal K}=\tilde{\sigma}_x(-\phi).
\end{eqnarray}

\section*{References}

\bibliographystyle{naturemag}

\begin{thebibliography}{99}
\bibitem[S3]{Raman}
We assume that the other states of an atom are energetically
separated from the three states by a large amount. 
Thus, the effective Hamiltonian becomes a simple form that includes 
only the three states.
\end{thebibliography}

\bigskip

\acknowledgments
We thank T. Fukui for providing us inspiration for this study.
Y. K. acknowledges the support of a Grant-in-Aid for JSPS
Fellows (No.17J00486). 
This work was partially supported 
by a Grant-in-Aid for Scientific Research from the Japan Society for the 
Promotion of Science under Grant Nos. 25220711, 16H00990, 
16H00801, and 17H06138,  
the Impulsing Paradigm Change through Disruptive Technologies (ImPACT) program, 
and JST CREST(No. JPMJCR1673).


\begin{center}
\textbf{\Large Supplementary Information}
\end{center}

\makeatletter
\setcounter{equation}{0} \setcounter{figure}{0}
\setcounter{table}{0} \setcounter{page}{1} \makeatletter
\renewcommand{\theequation}{S\arabic{equation}}


\section*{$\Lambda$-shaped schemes} 
Following Refs.\cite{Fox,KeilmannA}, 
we shall briefly explain the basic structure of laser-assisted hopping, 
which we call the $\Lambda$-shaped scheme.
Two ground states of neighboring sites, $|A\rangle$ and $|B\rangle$ 
and one excited state, $|E\rangle$, which is energetically higher than 
the two ground states, are used. 
The three level Hamiltonian $H_{\rm ABE}$ is given by \cite{Raman}
\begin{eqnarray}
H_{\rm ABE}=\sum_{l=A,B,E}\epsilon_{l}|l\rangle\langle l|+\frac{\Omega_{AE}}{2}|e\rangle\langle g_{j}|
+\frac{\Omega_{BE}}{2}|e\rangle\langle g_{j+1}|+\mbox{h.c.},
\end{eqnarray}
where $\epsilon_{l}$ is the energy for each state. 
The two Raman lasers have the wave-vectors ${\bf k}_{A}$ and ${\bf k}_{B}$.
Here, we assume that the diagonal terms of $H_{\rm ABE}$ are 
sufficiently larger than 
the off-diagonal terms and that the detuning is much smaller 
than $|\Omega_{AE}|$ and $|\Omega_{BE}|$.
Then, we use second-order perturbation theory 
with the rotating wave approximation (RWA) \cite{Fox,KeilmannA}. 
The effective transitions between the two adjacent ground states are given as 
\begin{eqnarray}
H^{(g)}_{\rm eff}&=&-\frac{1}{4\delta}
\begin{bmatrix}
  |\Omega'_{AE}|^{2} & \Omega'^{*}_{BE}\Omega'_{AE} \\
  \Omega'^{*}_{AE}\Omega'_{BE} &  |\Omega'_{BE}|^{2}
\end{bmatrix},
\end{eqnarray}
where $\Omega'_{AE(BE)}$ is the rotating frame representation. 
The diagonal terms are just the effective energy shifts. 
If $|\Omega'_{AE}|=|\Omega'_{BE}|$, they are negligible, 
as they just gives a uniform energy shift for each lattice site.

\section*{References}
\bigskip

\end{document}